\documentclass[journal=jacsat,manuscript=article]{achemso}
\usepackage[T1]{fontenc} 

\usepackage{achemso}
\usepackage[final]{pdfpages}
\usepackage[breaklinks=true,colorlinks=true,linkcolor=blue,urlcolor=blue,citecolor=blue]{hyperref}
\usepackage{amsmath,amssymb}
\usepackage{mathtools}
\usepackage{cleveref}
\usepackage{flexisym}
\usepackage{braket}     
\usepackage{graphicx}   
\usepackage{verbatim}   
\usepackage{color}      
\usepackage{subfigure}  
\usepackage{hyperref}   
\usepackage[utf8]{inputenc}
\usepackage[english]{babel}

\author{Petr Lega}
\affiliation{The Kotel'nikov Institute of Radio Engineering and Electronics, Russian Academy of Sciences, Moscow, 125009, Russia}
\email{lega_peter@list.ru}

\author{Alexey Karstev}
\affiliation{Computing Center FEB RAS, Khabarovsk, Russia.}
\alsoaffiliation{School of Mathematics and Physics Queen's University Belfast Belfast BT7 1NN Northern Ireland United Kingdom.}
\email{karec1@gmail.com}

\author{Ilya Nedospasov}
\affiliation{The Kotel'nikov Institute of Radio Engineering and Electronics, Russian Academy of Sciences, Moscow, 125009, Russia}

\author{Shuhui Lv}
\affiliation{School of Materials Science and Engineering, Changchun University of Science and Technology, Changchun 130022, China}

\author{Xiaoling Lv}
\affiliation{School of Materials Science and Engineering, Changchun University of Science and Technology, Changchun 130022, China}

\author{Natalia Tabachkova}
\affiliation{National University of Science and Technology MISiS, Moscow, Russia}

\author{Artemy Irzhak}
\affiliation{National University of Science and Technology MISiS, Moscow, Russia}
\alsoaffiliation{Institute of Microelectronics Technology and High-Purity Materials, Russian Academy of Sciences, Chernogolovka, Moscow oblast, Russia}

\author{Andrey Orlov}
\affiliation{The Kotel'nikov Institute of Radio Engineering and Electronics, Russian Academy of Sciences, Moscow, 125009, Russia}
\alsoaffiliation{Institute of Nanotechnology of Microelectronics, Russian Academy of Sciences, Moscow, 115487, Russia}

\author{Victor Koledov}
\affiliation{The Kotel'nikov Institute of Radio Engineering and Electronics, Russian Academy of Sciences, Moscow, 125009, Russia}

\title{Blocking of martensitic transition at the nano-scale in the Ti2NiCu wedge.}
\date{\today}

\begin{document}

	\maketitle

\begin{abstract}
	Shape memory effect associated with martensitic transformations is of the rapidly developing field in nanotechnologies, where industrial use of systems established on that effect provide greater flexibility on the nano-devices fabrication of various kind. And therefore it addresses questions to the phase transition phenomena at low-scale and its limitations and control. In this report, we studied the crystal structure of tapered plates of Ti$_2$NiCu alloy and the temperature $T_c$ at which the martensitic transition occurs. We demonstrated that $T_c$ has a strongly descending character as a function of the plate thickness $h$. The critical thickness value at which the transition completely suppressed is 20 nm. Moreover, the obtained results for $T_c(h)$ curves indicate the hysteretic nature of the transition. These findings open the pathway for size limits indication and regimes modulation where the alloy-based nano-mechanical devices can be tuned to operate more efficiently.
\end{abstract}

\footnotetext{\dag~Electronic Supplementary Information (ESI) available: See DOI: 10.1039/b000000x/}



%
\section{Introduction}
One of the most challenging as well as fundamental limitations associated with condensed matter physics is to assess physical properties of different materials at the nano-scale. For instance, the quantum-dimensional effect, causing a change in the thermodynamic and kinetic properties of thin metallic films and nano-particles ~\cite{se1,se2,se3,se4,se5,se6}, takes place in systems, where at least one of their dimensions is in the range of de Broglie wavelength. However, the problem is far deeper. Compared to phase transformations in massive systems such as bulk crystallines, size-dependent phase transitions developing in samples at the micro- and nano-level can play a significant role in their physicochemical nature and it's understanding. For instance, the melting temperature of gold particles strongly depends on their size and environment~\cite{Ag_prtcls2,Ag_prtcls3,Ag_prtcls4,Ag_prtcls5,Ag_prtcls6}, with their size reaching critical number 2.5 nm~\cite{Ag_prtcls1} at room temperature. 

Not only liquid-to-solid but also solid-to-solid phase transformations involve dimensional changes.  For example, it is essential for quasi-one-dimensional conductors with a charge density wave of cross sizes considerably lower than 1 $\mu m$ where many properties are influenced by the size effect~\cite{1D_CDW1,1D_CDW2}. Among phase transformations in intermetallic compounds, thermoelastic martensitic transition from a cubic high-symmetry phase (austenite) to a low-symmetry phase (martensite)~\cite{martensite1,martensite2,martensite3,martensite4} has attracted much interest these days. Here central to the entire discipline is the concept of shape memory effect (SME). SME is that a sample returning to its initial shape upon heating above the martensitic temperature as it been observed in some materials. SME finds presently a broader application in instrument-making, medicine, micro- and nano-mechanics ~\cite{SME}, where micromechanical devices of smallest sizes have been fabricated and exploited successfully~\cite{device1, device2, device3}.

\begin{figure*}
	\centering
	\includegraphics[width=0.77\paperwidth]{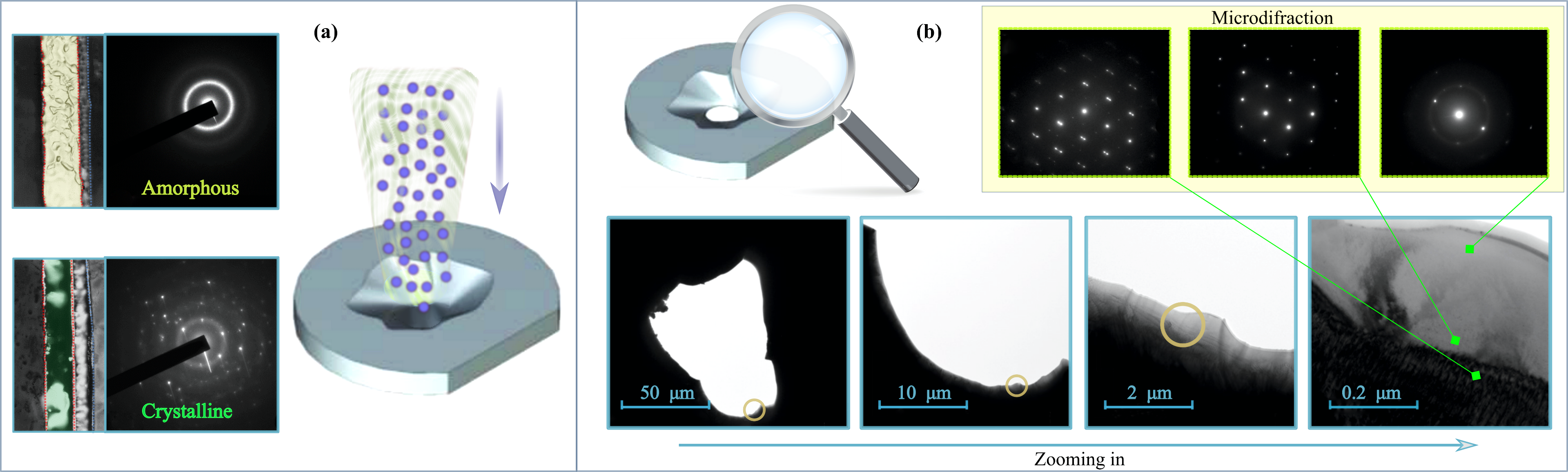}
	\caption{\label{figure:hole} \textbf{(a)} Microphotographs and microdiffraction of Ti$_2$NiCu ribbons in the amorphous state and annealed ribbon in the crystalline state. Yellow and green colours indicate amorphous and crystalline bulk phases correspondently. The later was thinned using the focused ion beam method. \textbf{(b)} High-resolution microphotographs at low-temperature of the wedge local areas in the imaging and in diffraction mode.}
\end{figure*}

At present, a key issue in the field is the determination of physical and technological limits as to the minimum size of the device that can function based on SME. To iron out this obstacle, it is necessary to study the effect of thermoelastic martensitic transition in samples of submicron size and deformation of micro- and nano-size samples under the influence of temperature and external mechanical stresses. This can be rationalized to the problem of the critical size of a particle in which the martensitic transition occurs. And it is similar to the classical problem of determining the critical size of the nucleus in a first-order phase transition~\cite{Landauv5}. Existing research on the nucleation of the martensitic phase recognises the important role of the grain size: if it decreases up to the nanometers size, the transition temperature of the alloy decrease in comparison with the bulk sample. With a further decrease in the grain size, the phase transition ceases ~\cite{kajiwara,kexel}.

The critical role played by the particle size in the martensite properties is admitted likewise~\cite{sizeeffect1,sizeeffect2,sizeeffect3}. In the work~\cite{glezer}, it was shown that the B2 $ \Rightarrow $  B19 thermoelastic martensitic transition arising in the Ti$_2$NiCu nanoparticles is blocked at the minimum nanoparticle size $ \lesssim $ 15 nm, while the non-thermoelastic martensitic transition $ \gamma $ $ \Rightarrow $ $ \alpha $ occurring within the Fe-Ni-B nanoparticles is suppressed at the greater nanocrystals size of the 100 nm order. A proposed theoretical model consider the energetic dispute between the volume free energy of favourable phase and the surface energy and the corresponding size factor, in analogous to the theory of rubber-like behaviour in Cu-nanowires~\cite{cunanowires}. This approach allowed to estimate a critical radius. However, attempts to calculate the temperature dependence of the critical radius or consider another geometry, for example, the emergence of a new phase in a flat film or on the edge of a wedge, never been made. 

A similar effect of the transformation temperature decrease been observed at the boundary between the amorphous and crystalline regions~\cite{santamarta}. These akin to that occurs in nano-spherulites, where the martensitic transformations are suppressed. Nevertheless, the authors did not attempt to cool down the sample and study the martensitic transformations at lower temperatures. In the same time, a non-monotonic behaviour for the temperature dependence on the thickness been observed for B2-R-B19$^\prime$ transformation in a wedge-shaped NiTi plate~\cite{pan2014thickness} where the important role plays the transition phase R. Whereupon another yet mechanism related to the surface oxidation been proposed for a NiTi alloy exhibiting complete suppression of the martensitic transformation~\cite{Li}. As can be seen from above the exact mechanism and its fundamental aspects that underpin transformation blocking are not fully understood. And, neither the actual limiting values for effects nor reliable theoretical explanations have been obtained yet. This study, therefore, set out to an experimental investigation into the thermoelastic martensitic transition in wedge-shaped plates of the Ti$_2$NiCu alloy. Along with the experimental analysis, the second aim is to performer theoretical investigation with the use of an adequate physical model explaining experimental results for the thickness and temperature in the range of 10-100 nm and 100-400 K correspondently.

\begin{figure}
	\centering
	\includegraphics[width=0.4\paperwidth]{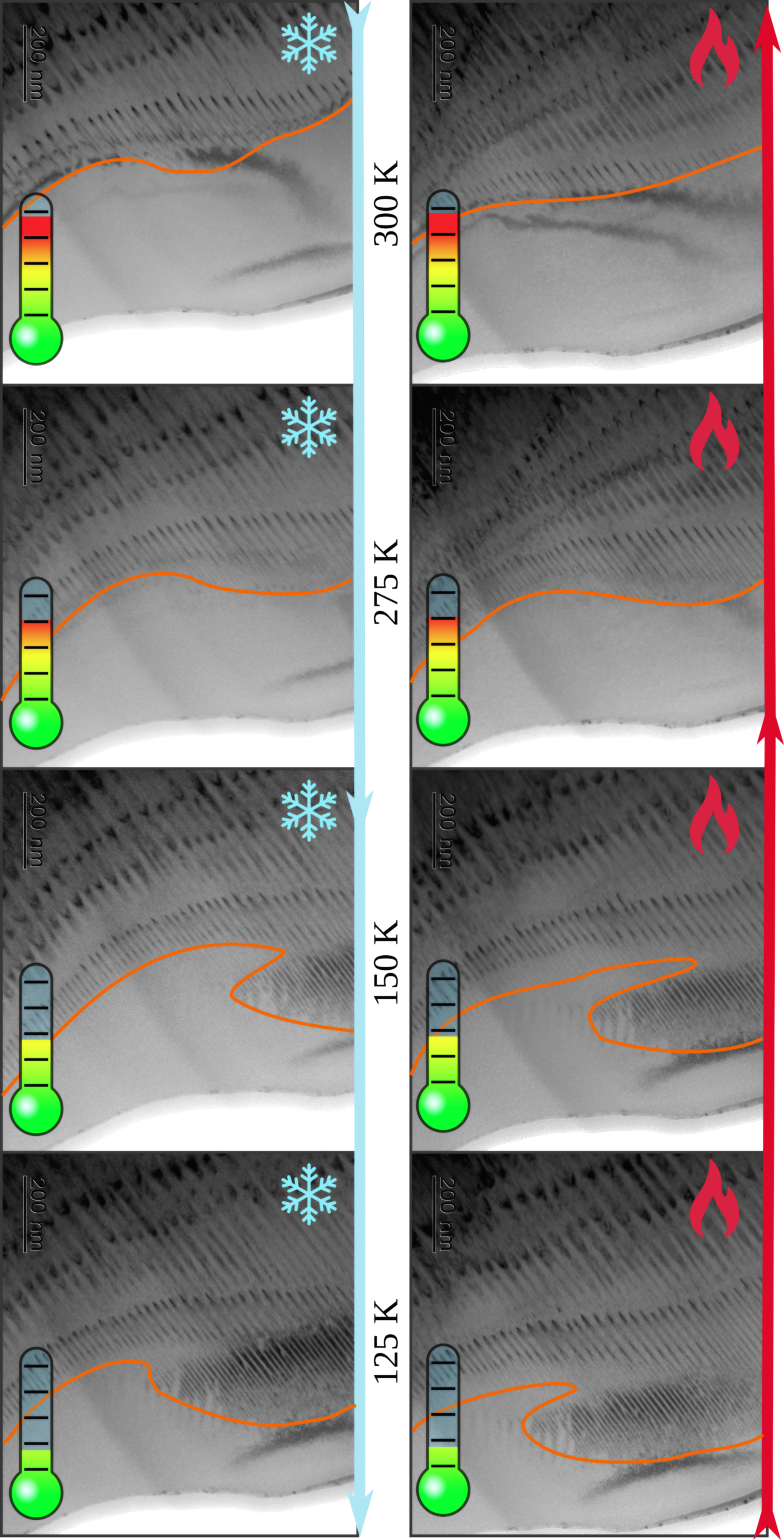}
	\caption{\label{figure:front} Microphotographs obtained using TEM in the fixed area of Ti$_2$NiCu wedge sample, near the edge of the sample at different temperatures. The transition boundary between phases is clearly visible and indicated by the red-orange line. The blue arrow corresponds to the cooling process (snowflake symbol), while the red arrow indicates temperature change during the heating process (flame symbol).}
\end{figure}

\begin{figure*}
	\centering
	\includegraphics[width=\columnwidth]{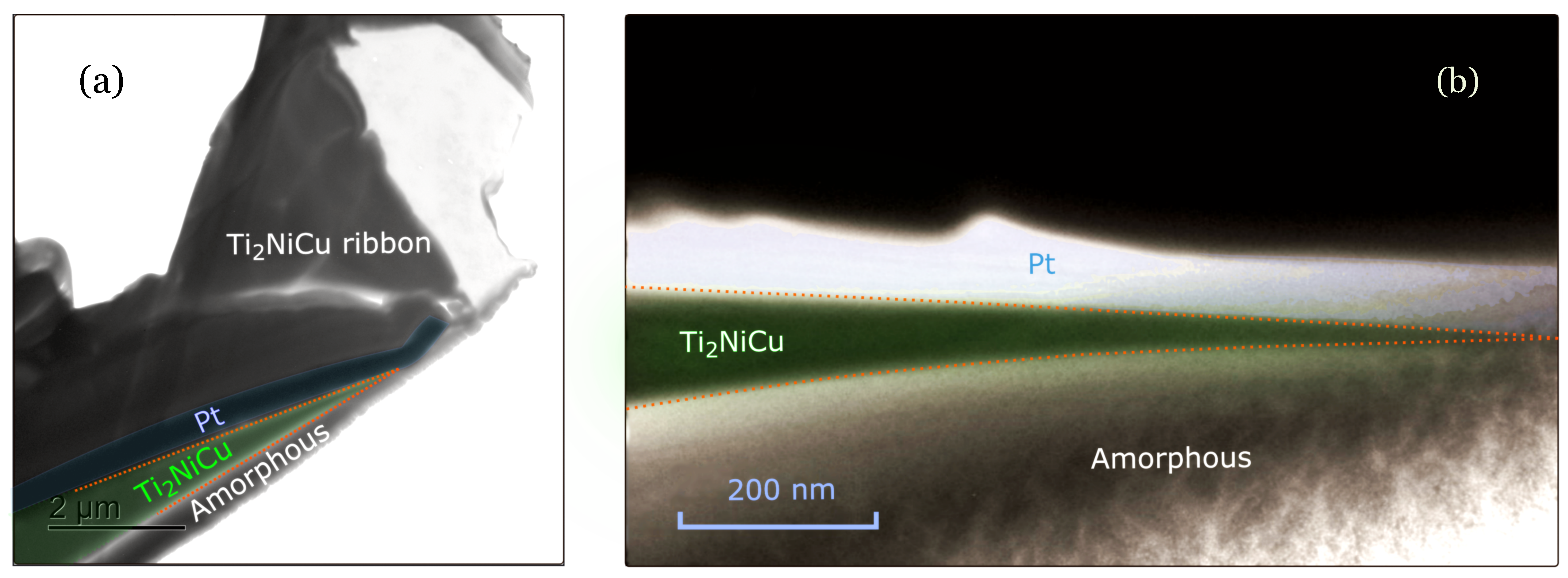}
	\caption{\label{figure:wedge}\textbf{(a)} TEM microphotograph of the studied Ti$_2$NiCu ribbon where wedge-shaped plate indicated by the light-green colour. The wedge boundaries indicted by the red-dashed line next to the platinum substrate (bluish colour) and amorphous phase. \textbf{(b)} The local area of the wedge-shaped part of a sample made using the focused ion beam (FIB) method and loaded into the TEM to examine its cross-section. To prevent the destruction of the surface by an ion beam, the sample was initially coated with a protective film of platinum of 400 nm thickness. The dark area in the upper part of the photo above the wedge-shaped sample is a protective film of platinum.}
\end{figure*}

\section{Results and Discussion.}
Initially, the Ti$_2$NiCu alloy was produced in the form of an amorphous ribbon and examined with the use of TEM. After annealing it to the crystalline state, the alloy underwent a thermoelastic martensitic transition from the cubic austenitic phase with the B2 structure to the orthorhombic martensitic phase with the B19 structure (Fig.~\ref{figure:hole} (a)). Subsequently, a sample of rapidly quenched ribbon was thinned using the ion thinner by the focused ion beam method (FIBM) until a hole with wedge-shaped edges appeared. The next step was to use a high-resolution TEM along with thermal stabilizer, where we studied the crystal structure of individual local areas in the imaging and in diffraction mode. It has demonstrated the local region near the edge is precisely in the austenitic phase. Microdiffraction patterns unequivocally prove that part of the region near the plate edge is indeed austenitic, despite the fact that austenite in a bulk alloy sample disappears at temperatures well below $M_f = 337$ K. Moreover, near the edge, the transition does not occur even at $T = 100$ K. Since the reaction/transformation front movement stops even earlier, at $T = 150$ K (Fig.~\ref{figure:hole} (b)).

TEM images obtained in the imaging mode  (see Fig.~\ref{figure:front}) allowed us to reveal the evolution of the regions, occupied by the martensitic and austenitic phases on the edge of the plate. Transition boundaries were clearly visible on all images, except the case of a bulk sample when it was filmed at the temperature above A$_f= 209$ K - no microtwins been observed due to the absence of the martensitic phase at high temperatures. Fig.~\ref{figure:front} shows a set of microphotographs obtained from the same area that is closer to the edge of the sample using the TEM at different temperatures. Where the phase boundary is distinguishable by the characteristic pattern of twins in the martensitic phase. It clear that the boundary moves with decreasing temperature, approaching the edge, that is, closer to the region of the minimum thickness of the plate. 

One can note that during the heating (see Fig.~\ref{figure:front}), the movement takes place at a higher temperature. And while the reaction front of the martensitic transition stops to move at $T = 150$ K on the cooling, there is no propagation during the heating process up to the $T = 220$ K. This pattern can be understood due to the fact that the first-order martensitic transition shows a hysteresis character. If the transformation has been reversed, i.e. when the sample been cooled the reaction front/boundary does not start to move immediately on the following heating. Front manoeuvres only after temperature reaching A$_f$, where some overheating occur. From the energy point of view, this is necessary for the system in order to overcome the potential barrier - to achieve the stability loss of an unfavourable low-temperature martensitic phase.

\begin{figure}[h!]
	\centering
	\includegraphics[width=0.7\columnwidth]{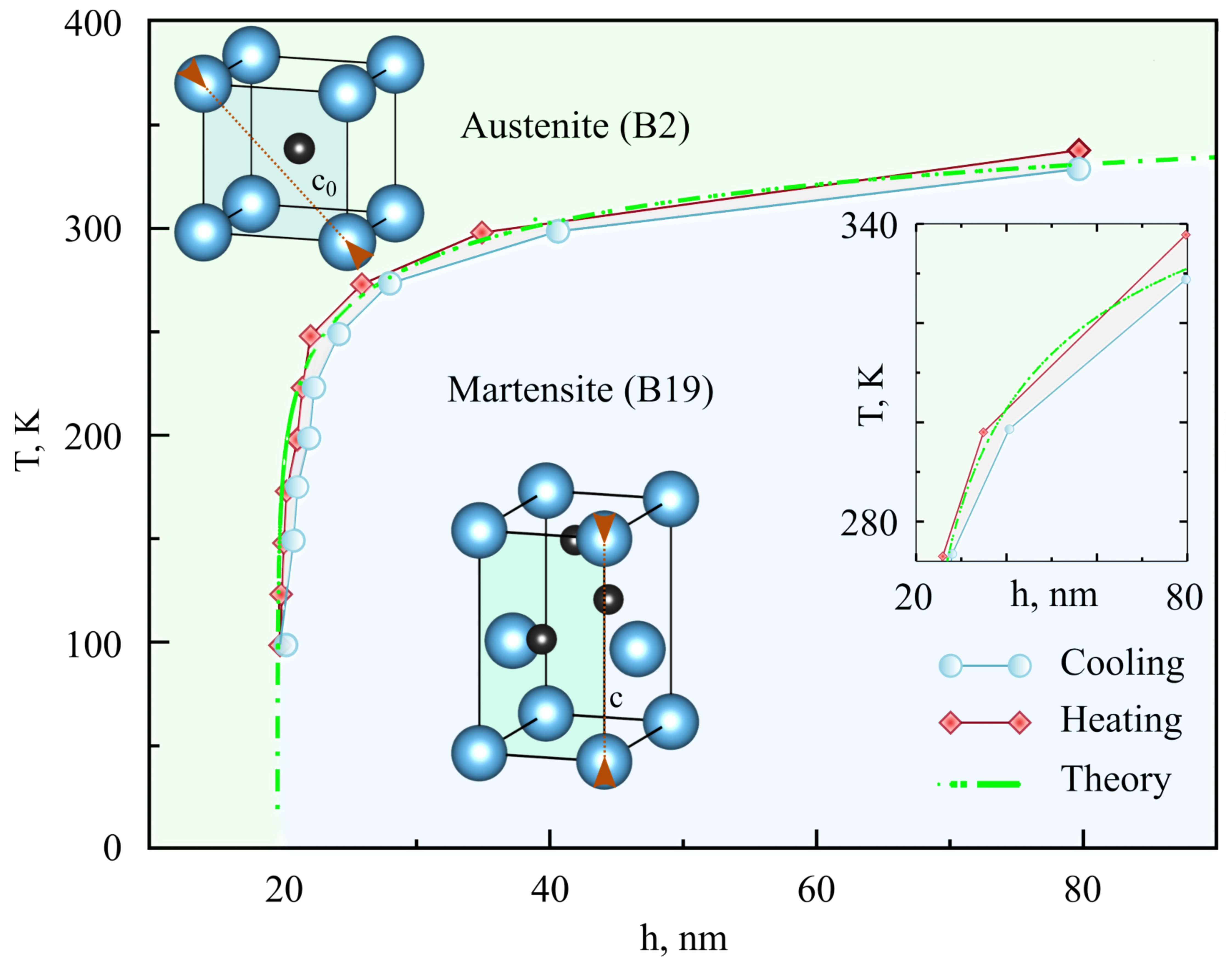}
	\caption{\label{figure:Th} Dependence of the martensitic transformation temperature on the plate thickness both for cooling and heating processes. Schematic representation of embedded nanocrystal of martensitic phase into a parent austenitic phase. Unit cell of B2 and B19 phases where red arows indicates correspondance in the lattice parametrs between two phases. The dashed green line represents the result of the dislocation-kinetic approach. The red-orange arrows indicate the relation between lattice parameters for the parent and daughter phases according to the Bain transformation. The insert indicates a zoomed region of the graph between 280 and 340 K where the main part of hysteresis is situated.}
\end{figure}

By means of the TEM measurements of the cross section in a sample, we have obtained the dependence of the plate thickness on the distance to the edge. Figure~\ref{figure:wedge}(a) illustrates the fabricated sample loaded into the transmission electron microscope (TEM) to examine its cross-section and to correlate the thickness of the wedge-shaped plate (Fig.~\ref{figure:wedge}(b)). Mobilizing the entire data set, we plotted the martensitic transformation temperature as the function of the plate thickness, as shown in Fig.~\ref{figure:Th}. What can be clearly seen in this figure is the critical character of the transition temperature dependence. With a shrink in the thickness from the 80 nm down to the 20 nm, the transition temperature decreases down to the 150 K and then drops sharply. Further cooling does not provoke any expansion of the region of the martensitic phase and some part of the sample remains in the austenitic phase thereof. And hence it will not lead to the formation of a low-temperature phase in these areas.

The most striking observation to emerge from the cooling and heating $T(h)$ data comparison was a hysteresis. The same area on the edge of the sample, where austenite state been inhabited at room temperature, transforms to the martensite on cooling at a certain temperature $T_1$. And proceeded by heating, it returns to the austenitic state at a bigger temperature $T_2>T_1$. This hysteresis-like character obviously can be attributed to the intrinsic properties of the martensitic first-order phase transition and its nanoscale nature.

\begin{figure}[h!]
	\centering
	\includegraphics[width=0.8\columnwidth]{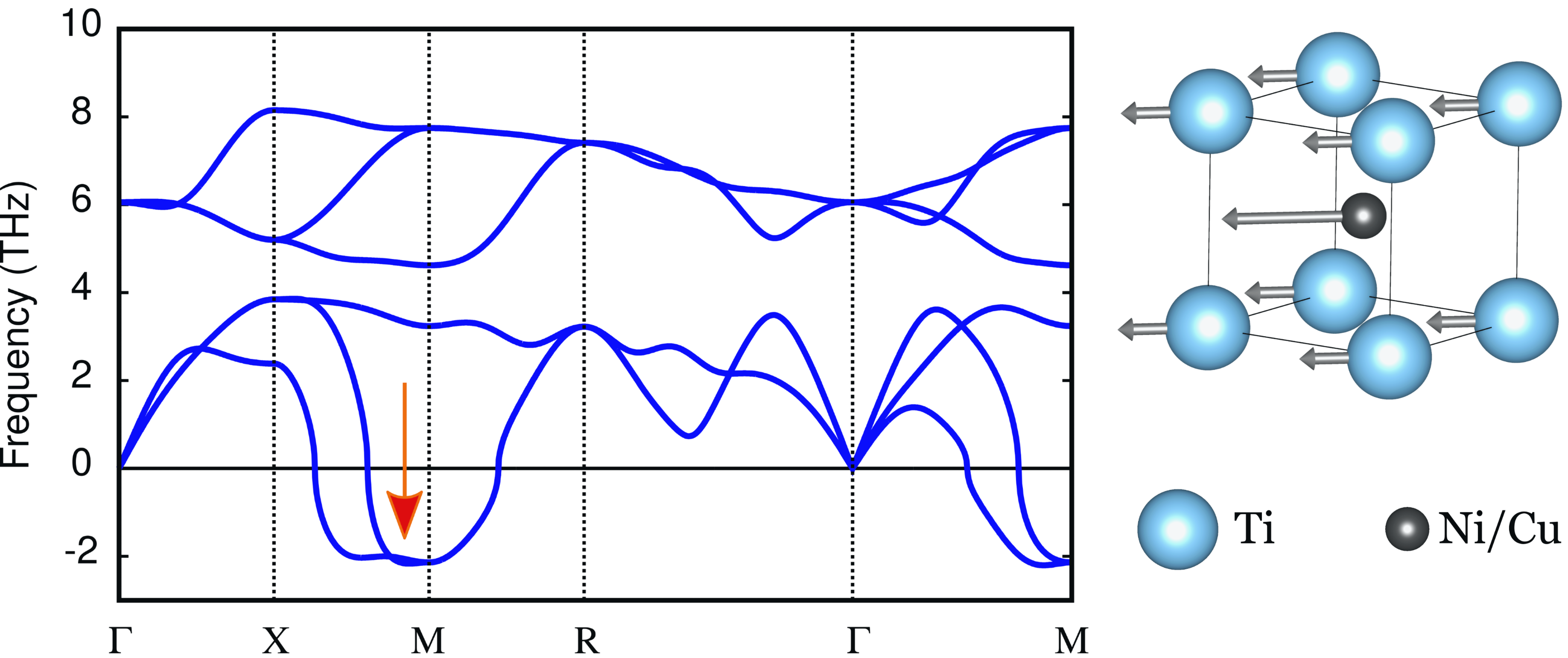}
	\caption{\label{figure:ph} Calculated phonon spectra of the B2 Ti$_2$NiCu phase using DFPT method. Imaginary frequencies of unstable modes are plotted as negative values, and the mode responsible for the B2 to the B19 transition is marked by the red arrow at the bottom. The corresponding acoustic phonons mode is shown on the right side where light-grey arrows indicate normalized atoms displacements in the B2 unit cell.}
\end{figure}

On the macroscopic level, this transition can be associated with the phonon softening during the martensitic transformation in the B2 phase~\cite{softph}. The phonon spectra at $T=0$ K of the Ti$_2$NiCu austenitic phase been calculated using DFPT and shown on Fig.~\ref{figure:ph}. The soft mode at the M point of the first Briliuon zone indicates the dynamic instability of the B2 phase. Where atomic displacements in the [110] direction correspond to martensitic transformation and define in-habit-plane \{110\} with a shear instability accompanied by the lattice distortion. The electronic nature of this softening originates from the to charge density wave instability related to nesting vectors in the Fermi surface.~\cite{Fnesting1,Fnesting2,Fnesting3} (see supplement materials). It is obvious that phonon dispersion curves of the parent cubic phase contain information closely related to the potential martensite phases and related transformations. This is essential for the identifying of B2-B19 transformation in the Ti$_2$NiCu system as the martensitic one. Where suppressing of the soft mode due to the finite size effect can lead to the metastable nature of austenite.


%
From the general thermodynamic point of view, the observed behaviour can be understood as a balance between different energetic impacts to the Gibbs energy. Consider the steady state of the system the following relation can be written~\cite{thermodyn} for the austenite to martensite transformation as a function of temperature $T$ and the sample thickness $h$:
\begin{equation}
G^{A \rightarrow M}(h,T)=\Delta G^{bulk}+ \Delta E^{surface} + \Delta E^{interface},
\end{equation}
where $\Delta G^{bulk}$ is the free energies difference for bulk phases, $\Delta E^{surface}$ the difference between surface energies and the $\Delta E^{interface}$ is the energy consuming to create new interfaces such as the boundary between phases/coating and the coherent twin boundaries.

While in the bulk $\Delta G^{bulk}$ is the main driving force for martensitic transformation, in the low-scale system it is primarily counteracted by the $\Delta G^{surface}$ and $\Delta G^{surface}$. We calculated the surface energies of Ti$_2$NiCu with both cubic and orthorhombic structures. The calculated values bear witness that in the bulk Ti$_2$NiCu martensite is more stable than the austenite at low temperatures. However, the Ni/Cu-terminated austenite Ti$_2$NiCu is the most stable surface structures. Therefore at the low temperature limit $\Delta G^{A \rightarrow M} (h=\infty, T=0) < 0$ and $\Delta G^{A \rightarrow M} (h=0, T=0) > 0$. While $\Delta G^{A \rightarrow M} (h=\infty, T> M_f) > 0$ reflecting the fact that in bulk the austenite is the high temperature phase. The thicker plate likely facilitates the creation twin boundaries and interfaces with a small contacting area, which then be compensated by the transformation strain across the different variants in order maintain the structural integrity of the wedge~\cite{HfOdomains,nanocrystals}. With the decrease of $h$, the surface energy will gain a major role and eventually may overcome $\Delta G^{bulk}$ leading to the austenite stabilization. While the opposite situation proceeds for the temperature decrease  -- $\Delta G^{bulk}$ increase and the martensite become more preferable. All the above arguments clearly indicate the importance of the interplay between surface and bulk energies in a wedge, where $\Delta G^{A \rightarrow M}(h, T)$ can be modulated via $h$ regulation and temperature regimes.

However, at the nano-scale, the finite-size effect is not visible for the heterogeneous thermodynamic treatment, hence it is deserved to describe the transformation blocking based on its microscopic nature. We applied another approach for the transformation description -- the analysis of interfacial dislocation movements where finite size effects can emerge. Using the well-established kinetic theory of dislocation as a martensite formation mechanism we were able to reach an agreement with experimentally observed $T(h)$ values (see Fig.~\ref{figure:Th}), where $T(h)$ defines by the following relation (see Appendices~\ref{sssec:DKT}):
\begin{equation}
T(h) \simeq T_{c} \left[1+B\ln \left(\frac{2k_{0} }{(1-k_{a} /3)+\lambda /h} -1\right)\right].
\end{equation}

\section{Methods}

\subsection{Sample and Experimental Details}

For the present study the Ti$_2$NiCu alloy was obtained in the amorphous state in the form of ribbons of 40 $ \mu $m thickness by ultrafast quenching of the molten alloy onto rotating copper drum ~\cite{matveeva1997effect}. The amorphous alloy was annealed in a furnace at 500$ ^{\circ} $  C for 4-6 min. In the crystalline state, the alloy undergoes a thermoelastic martensitic transition from the cubic austenitic phase with a B2 structure to the orthorhombic martensitic phase with a B19 structure. The temperatures for the start and finish for the forward and reverse martensitic transitions of the samples used in this work are, respectively: M$_s$ = 60$ ^{\circ} $ C, $M_f$ = 52$ ^{\circ} $ C, A$_s =$ 55$ ^{\circ} $ C, and A$_f$ = 64$^{\circ}$ C.

In the process of preparing the samples for the experiments, the samples of rapidly quenched ribbon were initially thinned using a GATAN Model 691 ion thinner (JEOL, Japan) until a hole with wedge-shaped edges appeared. Then, using a high-resolution TEM JEM-2100 (JEOL, Japan) with a GATAN thermal stabilizer, we studied the crystal structure of individual local areas in the imaging and in diffraction mode in the temperature range of 100-400 K. Images obtained using the TEM in the imaging mode allow us to reveal the evolution of the regions, occupied by the martensitic and austenitic phases, on the edge of the plate. The phase boundary is distinguishable by the characteristic pattern of twins in the martensitic phase. The current density on the sample was applies in a control way. Where under these experimental conditions, it did not exceed 1 mA/cm$^2$, which can lead to a negligible small increase in the temperature of the sample ($< 1^{\circ}C$).

To determine the thickness of the wedge-shaped plate in the TEM study areas, a cross section of the sample was made using the focused ion beam (FIB) method. To prevent the destruction of the surface by an ion beam, the sample was initially coated with a protective film of platinum of 400 nm thickness. The sample obtained was loaded into the TEM to examine its cross-section and to correlate the thickness of the wedge-shaped plate to the distance to the edge. The structural conclusions of the results obtained from TEM were confirmed by microdiffraction patterns in local regions of a sample. Additionally, in order to verify observed numbers the thickness measurement been performed by employing the electron energy loss spectroscopy (EELS) where obtained results were in the correspondence with TEM measurements.

\subsection{Computational details}

The geometry optimization, total energies and electronic structures calculation were performed using the Vienna \textit{ab initio} simulation package (VASP) under the framework of density-functional theory (DFT)~\cite{vasp1,vasp2}. The projector-augmented wave (PAW) method~\cite{paw} was used for the electron-ion interactions and the generalized gradient approximation (GGA) of Perdew, Burke and Ernzerhof (GGA-PBE) ~\cite{GGA} was employed to describe the exchange-correlation function. A cutoff energy of 500 eV was used and the irreducible Brillouin zone was sampled with a regular Monkhorst-Pack grid~\cite{MP} of 5$ \times $ 5$ \times $ 5 and 4$ \times $ 5$ \times $ 5 \textit{k}-points for the total energy calculation of bulk cubic and orthorhombic Ti$_2$NiCu respectively, and 4$ \times $ 4$ \times $ 1 and 5$ \times $ 5$ \times $ 1 \textit{k}-points for the surface calculation of cubic and orthorhombic Ti$_2$NiCu, respectively. For the surface calculations, a vacuum region of 10 Å was added to avoid the unwanted interaction between slabs and its period images. All atoms were fully relaxed until the magnitude of forces on each atom converged to less than 0.05 eV/Å. 

For the surface calculation in the framework of the super-cell method for cubic Ti$_2$NiCu, two types of (001)-oriented surfaces are considered where occurs the most atomic rearrangement during the martensitic transformation. That are NiCu and Ti terminations. For the surface calculation of orthorhombic Ti$_2$NiCu, we considered four types of (001) direction termination surfaces, i.e., NiCu-1, NiCu-2, Ti-1 and Ti-2 for Ti$_2$NiCu (Supplementary Fig. 1). To ensure that the two sides of the surface slabs used in the calculation are thick enough to exhibit bulk-like interiors, we made calculations of the surface energy with respect to sufficiently thick slabs and then performed full surface relaxations. Surface energies of the Ti$_2$NiCu under the condition that the chemical potential of the respective element equals its bulk total energy can be read as follows:
\begin{equation}
\begin{split}
\sigma_{Ti_2NiCu} &=\frac{1}{2A}\left[ E^{total}_{slab}-\frac{1}{2}N_{Ti}\cdot E^{bulk}_{Ti_2NiCu}\right. \\
& \left. -E^{bulk}_{Ni}\left(N_{Ni}-\frac{1}{2}N_{Ti}\right)-E^{bulk}_{Cu}\left(N_{Cu}-\frac{1}{2}N_{Ti}\right) \right],
\end{split}
\label{eq:surfenergy}
\end{equation}
where $E^{total}_{slab}$ and $E^{bulk}_{X}$ indicate total energies of a slab and \textit{X} bulk, and \textit{N}\textsubscript{Y} and \textit{A} represent the number of Y (Y = Ti, Ni, Cu) atoms and surface area, respectively. Using Eq. (\ref{eq:surfenergy}).

Finally, we also investigated the role of alloying by using the virtual crystal approximation (VCA) in the frame of a computational implementation of the density-functional theory (DFT) package Quantum Espresso~\cite{QE-2009}. The ground states calculations were performed using ultrasoft pseudopotential approach and the Perdew-Burke-Ernzerhof (PBE) generalized gradient approximation~\cite{PBE} of the exchange and correlation functional where the scalar relativistic corrections were included. We have employed a 600 eV energy cut-off and a 60 eV wave functions cut-off to optimize the ground state. The $15\times 15\times 1$ for a single unit cell and $8\times 8\times 1$ for the $2\times 2$ supercells calculations Monkhorst and Pack $k$-point meshes~\cite{MP} were used for integration in the irreducible Brillouin zone by a special-points technique with broadening $\sigma=0.02$ Ry according to the Marzari-Vanderbilt cold smearing method~\cite{MV}. Thus, these meshes ensure convergence of total energy to less than $10^{-6}$ eV/atom. The enthalpies difference and optimised lattice parameters as a function of hydrostatic pressure have been calculated with the use of variable cell-shape relaxation method. Alloy vibrational properties are determined with the density-functional perturbation theory (DFPT), with vibrational spectra and the corresponding normal modes obtained from the first-principle interatomic force constants by using $6\times 6\times 6$ $q$-meshes in the first Brillouin zones within frequency convergence less than 0.5 cm$^{-1}$. Calculated equilibrium lattice parameters for Ti$_2$NiCu are $a_0 =$3.05 \AA \ for B2 phase, and $a=2.726$ \AA, $b=4.359$ \AA, $c=4.701$ \AA \ for B19 phase correspondingly. And they are in the good agreement with experimental values ~\cite{latparamexper1, latparamexper2, latparamexper3} and previous theoretical estimations~\cite{latparamther1, latparamther2}.

\subsection{Dislocation-kinetic approach}\label{sssec:DKT}

For the further discussion, a method similar to that proposed in the Ref.~\cite{malygin}, namely, the kinetic approach to the formation of martensitic structures is considered, i.e. the process of self-organization of nucleus volumes of transformation. The transformation is carried out due to the movement of the dislocation transformations in the form of martensitic steps of atomic sizes at the interphase boundaries. In the theory of martensitic transitions, in early works, the following relations for the volume fractions of martensitic ($\phi_M$) and austenitic ($\phi_A$) phases were established:
\begin{eqnarray}
	\phi _{{\rm M}} =\left(1+\exp \left(\frac{\Delta U}{kT} \right)\right)^{-1};\\
	\phi _{{\rm A}} =1-\phi _{{\rm M}},
\end{eqnarray}
where $\Delta U=\omega \Delta u$, $\omega $  -- is the nucleus volume of transformation, \textit{T} -- correspond to the temperature, $\Delta u$ - is the change in internal energy per unit volume of a crystal during a structural transition, which can be determined by the following ratio:
\begin{equation}
	\Delta u=q\frac{T-T_{c} }{T_{c} } -\xi \sigma,
\end{equation}
where \textit{q} -- transition heat, $\xi \ $- spontaneous shear deformation of the lattice associated with the transformation, $\sigma -\ $ shear stress at uniaxial crystal loading, \textit{k} -- Boltzmann constant.

On the other hand, based on the mean free path for dislocation $\lambda$ involved in the transformation, we can write the kinetic equation for the relative proportion of martensite $\phi$ as the following phenomenological expression:
\begin{equation}
	\tau \frac{\partial \varphi }{\partial t}=k_0\nu+k_m\varphi -k_a\varphi \left(1-\varphi \right)+{\lambda }^2_D\frac{{\partial }^2\varphi }{\partial x^2},
\end{equation}

where $\tau =\ \lambda /v\ $-- characteristic time, $\nu$ is the generation intensity of sources at finite temperature, $v\ $-- dislocation speed, $h$ - martensitic step height, $n_0-\ $bulk density of sources of transformation dislocations, $h_a,\ {\lambda }_m,\ \ {\lambda }_D$- are characteristic distances, respectively, for the annihilation of martensitic and austenitic steps, multiplication of dislocations and diffusion of dislocations during their interaction with lattice defects, \textit{x} -- coordinate in the direction perpendicular to the plane of the boundary of the two phases. 

Since we assume that the temperature changes adiabatically, the main interest for us is the analysis of static ($\frac{\partial \varphi }{\partial t}=0)\ $solutions of the kinetic equation. The analysis of this solution was done by Malygin in details~\cite{malygin,malygin2}. This paper aims to establish relationships between the coefficients for determining the critical temperature. As a consequence, it is possible to establish the dependence of temperature on the size of the grain and the thickness of the layer:
\begin{equation}
	T(d,h)=T_{c} \left[1-B\ln \left(\frac{2k_{0} }{(1-k_{a} /3)+\lambda /d+\lambda /h} -1\right)\right]^{-1},
\end{equation}
where the phase transition smearing defined by $B=\frac{\omega q}{k_B T_c}$ and  $k_B$ is the Boltzmann constant.

In our work we consider a different approximation $d \gg h$ that the thickness of the plate is much smaller than the size of the grain, as a result of which we can neglect the member in brackets:
\begin{equation}
	T(h)\approx T_{c} \left[1-B\ln \left(\frac{2k_{0} }{(1-k_{a} /3)+\lambda /h} -1\right)\right]^{-1}.
\end{equation}

Thus, it turns out that the nucleus volume of the transformation per unit dislocation length is $a\rho \lambda $, where $a$ -- lattice parameter. The boundary of the wedge are barriers to the dislocation of the transformation and at small sizes limits the length of their free path $\lambda$. The corresponding values of parameters critical temperature in the free bulk $T_c=310$ K, $k_a=1.0$, $k_0=0.22$, the nucleus volume of transformation $\omega=4100$ nm$^3$ and have been chosen to fit the $T(h)$ curve to the experimentally observed values. Where the T=0 limit defines the minimal possible transverse size $h_{min}\approx 20 $ nm. This value well agrees with the geometrical estimation of the smallest possible martensitic block size which is based on the coherency principle to the parent B2 phase (see supplement materials). The obtained value of the standard enthalpy of martensite formation $q =5.5\cdot 10^{4}$ J/kg complies with estimated values 5.3-6.0 J/kg of martensitic formation energy $E^{tot}(B2)-E^{tot}(B19)$ calculated by two different methods in the frame of DFT. Such a reconciliation of quantities indicates the reliability of this approach. 

\section{Conclusion}

We performed a versatile investigated the crystal structures in the local areas of the wedge-shaped plates of the Ti$_2$NiCu alloy in the thickness range of 10-100 nm in the temperature range of 100-400K, where martensite phase been detected at the room temperature for plates thickness more than 80 nm. We also observed that the austenite phase proliferates with a decrease in temperature while the transition temperature decreases with decreasing plate thickness and has a hysteretic character. At a temperature of 150 K and a thickness of 20 nm, a critical point been identified, where the martensitic transformation is completely blocked. Aforementioned behaviour is found to be in good agreement with theoretical macroscopical and thermodynamic concepts. The approach based on a size-control paves a new way of the shape memory effect modulation at nano-scale and fine tuning work parameters of nanodevices. It will likely enable a broad variety of micromechanical machinery in the nanomechanics and nanotechnologies based on Ti$_2$NiCu alloy.

\section{Supplement material}
Computational details of surface and bulk total energies calculations for Ti$_2$NiCu and TiNi alloys.\\
Model for Laplace pressure as a driving force for martensitic transition in a cone-shape nano-rode.\\
Fermi surface for austenic phase and corresponding nesting vectors.\\
Parasitic heating of sample in the TEM during measurements.\\
Details on the surface energy calculations based on the super-cell approach.

\section{Acknowledgements}

The experimental part of this study was performed under support of the RSF grant No. 17-19-01748. The theoretical part was funded by the support of the RFBR grant No. 18-29-11051. We are grateful to the UK Materials and Molecular Modelling Hub for computational resources, which is partially funded by EPSRC (EP/P020194/1), and the Shared Facility Center resources "Data Center of FEB RAS" (Khabarovsk)~\cite{Sorokin2017} where the high-performance computing and data processing was performed under support of the RFBR grant No. 18-29-11051.

\bibliography{refs}

\end{document}